\documentclass{aa}

\usepackage{graphics}

\def\hi{H{\sc i}}
\def\xp#1{$10^{#1}$}
\def\xxp#1{$\times 10^{#1}$}
\def\ergs{erg s$^{-1}$}
\def\msun{M$_{\odot}$}
\def\cms{cm$^{-2}$}
\def\cmc{cm$^{-3}$}
\def\kms{km~s$^{-1}$}

\begin{document}

\title{The presence and distribution of HI absorbing gas in
sub-galactic sized radio sources}

\author{Y. M. Pihlstr\"om\inst{1,2}
   \and J. E. Conway\inst{2}
   \and R. C. Vermeulen\inst{3}
         }

 \institute{National Radio Astronomy Observatory, P.O. Box O, Socorro,
    NM 87801, USA \and Onsala Space Observatory, S-439 92 Onsala,
    Sweden \and Netherlands Foundation for Research in Astronomy,
    Postbus 2, 7990 AA Dwingeloo, The Netherlands}

\offprints{ypihlstr@aoc.nrao.edu}
\date{Received / accepted }

 
\abstract{We consider the incidence of \hi\ absorption in
intrinsically small sub-galactic sized extragalactic sources selected
from sources classified as Gigahertz Peaked Spectrum (GPS) and Compact
Steep Spectrum (CSS) sources. We find that the smaller sources
($<$~0.5 kpc) have larger \hi\ column densities than the larger
sources ($>$~0.5 kpc). Both a spherical and an axi-symmetric gas
distribution, with a radial power law density profile, can be used to
explain this anti-correlation between projected linear size and \hi\
column density. Since most detections occur in objects classified as
galaxies, we argue that if the unified schemes apply to GPS/CSSs a
disk distribution for the \hi\ is more likely. The most favoured
explanation for the compact sizes of the GPS/CSSs is that they are
young sources evolving in a power law density medium. For the GPSs
with measured expansion velocities, our derived densities are within
an order of magnitude of those estimated from ram-pressure confinement
of the lobes assuming equipartition. Our results therefore support the
youth model.

\keywords{Galaxies: active -- Galaxies: evolution -- Galaxies: ISM --
Radio lines: galaxies} }

\maketitle

\section{Introduction}\label{intro_stat}
Gigahertz Peaked Spectrum (GPS) and Compact Steep Spectrum (CSS) radio
sources have their radio emission confined to a very small region
($<$~10~kpc). Their radio spectra and radio morphology indicate this
class of sources are intrinsically small in contrast to the
relativistically beamed core-jet sources.

In these two classes of sources the turn-over frequency is
anti-correlated with the source size. This has led to the argument
that that the turnover could be due to synchrotron self-absorption
(Fanti et al.\ \cite{fanti90}; O'Dea \& Baum \cite{odea97}, Snellen et
al.\ \cite{snellen00}). A turn-over in the 100~MHz regime is
characteristic of the CSSs with sizes 1~--~10~kpc, while if the peak
flux density instead is seen close to 1~GHz, the sources are smaller
($<$~1~kpc) and are classified as GPSs. The most likely explanation
for the small sizes of these objects is that they are young sources
which will eventually evolve into larger scale classical doubles like
Cygnus A (e.g.\ Fanti et al.\ \cite{fanti95}; Readhead et al.\
\cite{readhead96}; Owsianik \& Conway \cite{owsianik98a}). They
therefore provide interesting laboratories for studying AGN evolution,
and present data is consistent with the idea that the sources expand
with constant velocities while their radio power declines as
$LS^{-0.5}$, where $LS$ is the linear size (e.g.\ Fanti et al.\
\cite{fanti95}). The drop of the radio power as the source grows in
size could indicate that the GPS sources may evolve into the lowest
luminosity FRII radio galaxies or perhaps into the FRIs (O'Dea \& Baum
\cite{odea97}).

The probable youth of GPS/CSSs implies the possibility of studying the
birth of radio sources; little is known of the mechanisms triggering
the radio activity. Mergers could provide a way to transport gas to
the centre of the host galaxies and thus be involved in the onset of
the nuclear engine. Indeed, optical observations have shown that many
of the GPS/CSS sources are in disturbed or interacting systems (de
Vries et al.\ \cite{devries00}; O'Dea et al.\ \cite{odea96}), in which
we may expect dense nuclear environments.

A number of studies of gas in GPS/CSS sources have already been made
at different wavebands. For instance, optical studies show strong
highly excited line emission with large equivalent widths (Gelderman
\& Whittle \cite{gelderman94}), consistent with interactions between
the radio source and the interstellar medium. Radio polarisation
observations have shown that some GPS/CSS sources display large
rotation measures exceeding $>$~1000~rad\,m$^{-2}$\ which in turn
imply large foreground electron densities and magnetic fields (Kato et
al.\ \cite{kato87}; Stanghellini et al.\
\cite{stanghellini98}). However, the majority of small GPS sources are
weakly polarised, making it hard to estimate rotation measures
(e.g. Stanghellini et al.\ \cite{stanghellini98}). The low
polarisation suggests strong depolarisation, again consistent with a
very large central density.  X-ray absorption observations of a few
GPS quasars indicate an absorbing column of a few \xp{22}~\cms\
associated with the quasars (Elvis et al.\ \cite{elvis94}). However,
there exist no extensive study of this phenomenon in the GPS/CSS class
as a whole.  Other evidence for a dense environment comes from
free-free absorption observations, for example in OQ208 (Kameno et
al.\ \cite{kameno00}). Marr et al.\ (\cite{marr01}) found evidence for
free-free absorption in the GPS source 0108+388, consistent with a
100~pc radius disk with an electron density of 500~\cmc. In 1946+708
multi-frequency continuum studies also show indications of free-free
absorption concentrated towards the core and inner parts of the
counter-jet, again suggesting a disk or torus origin (Peck et al.\
\cite{peck99}). Disk-like distributions of gas have also been found in
the optical, and the best example so far is the HST dust disk observed
in the GPS source 4C\,31.04 (Perlman et al.\ \cite{perlman01}).  Due
to the sparse number of observations, a typical molecular gas mass of
GPS/CSSs is not yet known, but CO observations of 4C\,12.50 have shown
a large total gas mass around \xp{10}\,\msun\ (Evans et al.\
\cite{evans99}; Mirabel et al.\ \cite{mirabel89b}). However this source
is unusual in many other of its properties and might not be
representative of the GPS/CSS class as a whole.

Another way to study the gas content in these sources is by spectral
absorption experiments. Such absorption observations have the advantage
over emission that the sensitivity is independent of the distance,
depending only on the background flux density. In absorption
experiments detection depends on having a significant column density,
rather than a large gas mass, so this method is sensitive to dense gas
on small scales in AGN. Using the 21 cm line of atomic hydrogen in
absorption it is possible to study the neutral gas content
of GPS/CSS sources. The \hi\ is likely to be only a fraction of the
total gas present, however such observations provide lower limits to
the total gas mass and density. The strength of GPS/CSS sources at cm
wavelengths makes them good targets for such experiments. In addition
their small sizes indicate that lines of sight to these sources will
sample the dense gas confined within the centre of the host
galaxy. Similarly the line of sight to a GPS source will trace gas
within the narrow-line region (NLR).

Here we discuss the results of \hi\ absorption studies in such
sources, compiled from our own observations presented in Vermeulen et
al.\ (\cite{vermeulen03}), and from the literature.


\section{The sample and its properties}\label{sample}

\begin{table*}
\caption[]{Cols.\ 1-3; source name, Cols.\ 4-11; optical redshift,
optical host galaxy ID, total observed \hi\ column density or a
3$\sigma$ upper limit, reference to the \hi\ absorption data, FWHM and
line peak opacity of the detected lines (if multiple lines are
present, the numbers given are for the component with highest column
density), linear size and finally total observed flux density at the
frequency of the redshifted 21cm line.}
\label{table1}
\begin{flushleft}
\begin{tabular}[t]{lllllrrrrrr}
\hline 
\hline\vspace*{-0.2cm}\\
Source&&Other&$z$&Id&log($N_{\rm HI}$)&\hi\,ref.&FWHM&$\tau _{\rm peak}$&Size&$S_{\rm cont}$\\ 
J2000 &B1950& name  & &&\cms& & \kms&10$^{-2}$&kpc  & Jy\vspace*{0.2cm}\\
\hline\vspace*{-0.2cm}\\
J0022+0015  &0019$-$000&4C\,00.02 &0.305  &G & ...    &...&... &...  &0.220 &...  \\
J0025$-$2602&0023$-$263&OB$-$238  &0.322  &G &20.4   &1  &126 &0.93 &2.429 &8.17 \\
J0111+3906  &0108+388  &OC\,314   &0.66847&G &21.9   &2  &94  &44.00&0.033 &0.18 \\
J0119+3210  &0116+319  &4C\,31.04 &0.060  &G &21.0   &3  &153 &3.70 &0.091 &2.60 \\
J0137+3309  &0134+329  &3C\,48    &0.367  &Q &19.3   &4  &100 &0.10 &1.124 &20   \\
J0141+1353  &0138+136  &3C\,49    &0.621  &G &20.0   &1  &35  &1.39 &5.001 &3.78 \\
J0224+2750  &0221+276  &3C\,67    &0.3102 &G &$<$20.1&1  &... &...  &9.585 &3.90 \\
J0348+3353  &0345+337  &3C\,93.1  &0.243  &G &$<$20.1&1  &... &...  &1.186 &2.95 \\
J0410+7656  &0403+768  &4C\,76.03 &0.5985 &G &20.4   &1  &61  &1.40 &0.689 &6.30 \\
J0431+2037  &0428+205  &OF\,247   &0.219  &G &20.5   &1  &297 &0.46 &0.653 &4.56 \\
J0503+0203  &0500+019  &OG\,003   &0.58457&Q &20.8   &2  &62  &3.60 &0.055 &1.60 \\
J0542+4951  &0538+498  &3C\,147   &0.545  &Q &$<$19.5&1  &... &...  &2.717 &28.75\\
J0556$-$0241&0554$-$026&...       &0.235  &G &$<$20.8&1  &... &... &$<$2.420&0.54\\
J0650+6001  &0646+601  &OH\,577.1 &0.455  &Q &...     &...&... &...  &0.014 &...  \\
J0713+4349  &0710+439  &OI\,417   &0.518  &Q &...     &...&... &...  &0.118 &...  \\
J0741+3112  &0738+313  &OI\,363   &0.635  &Q &$<$20.0&1  &... &...  &0.041 &1.33 \\
J0909+4253  &0906+430  &3C\,216   &0.670  &Q &20.1   &1  &177 &0.38 &9.349 &6.19 \\
J0943$-$0819&0941$-$080&...       &0.228  &G &$<$20.1&1  &... &...  &0.148 &3.19 \\
J1035+5628  &1031+567  &OL\,553   &0.45   &G &$<$20.1&1  &... &...  &0.163 &1.85 \\
J1111+1955  &1108+201  &OM\,214   &0.2991 &G &...     &...&... &...  &0.066 &...  \\ 
J1120+1420  &1117+146  &4C\,14.41 &0.362  &G &$<$19.8&1  &... &...  &0.306 &2.75 \\
J1124+1919  &1122+195  &3C\,258   &0.165  &G &...     &...&... &...  &0.242 &...  \\
J1206+6413  &1203+645  &3C\,268.3 &0.371  &G &20.3   &1  &101 &1.00 &5.641 &3.34 \\
J1252+5634  &1252+565  &3C\,277.1 &0.321  &Q &$<$19.8&1  &... &...  &6.146 &2.27 \\
J1308$-$0950&1306$-$095&OP$-$10   &0.464  &G &$<$20.1&1  &... &...  &1.710 &4.82 \\
J1326+3154  &1323+321  &4C\,32.44 &0.37   &G &19.9   &1  &229 &0.17 &0.247 &5.41 \\
J1347+1217  &1345+125  &4C\,12.50 &0.122  &G &20.6   &5  &150 &1.38 &0.166 &4.70 \\
J1357+4354  &1355+441  &...       &0.64   &G &21.5   &1  &367 &5.00 &0.067 &0.50 \\
J1400+6210  &1358+624  &4C\,62.22 &0.431  &G &20.3   &1  &170 &0.61 &0.218 &5.47 \\
J1407+2827  &1404+286  &OQ\,208   &0.07658&G &20.3   &1  &256 &0.39 &0.010 &0.79 \\
J1414+4554  &1412+461  &...       &0.186  &G &...     &...&... &...  &0.079 &...  \\
J1415+1320  &1413+135  &OQ\,122   &0.24671&Q &21.0   &6  &18  &34.00&0.120 &1.25 \\
J1443+7707  &1443+773  &3C\,303.1 &0.267  &G &$<$20.1&1  &... &...  &6.295 &2.12 \\
J1546+0026  &1543+005  &...       &0.55   &G &$<$20.0&1  &... &...  &0.049 &1.77 \\
J1609+2641  &1607+268  &OS\,211   &0.473  &G &...     &...&... &...  &0.223 &...  \\
J1643+1715  &1641+173  &3C\,346   &0.161  &G &...     &...&... &...  &8.834 &...  \\ 
J1815+6127  &1815+614  &...       &0.601  &Q &20.6   &1  &118 &2.03 &0.061 &0.97 \\
J1816+3457  &1814+349  &...       &0.2448 &G &20.7   &7  &80  &3.50 &0.123 &0.80 \\
J1823+7938  &1826+796  &...       &0.224  &G &$<$21.4&1  &... &...  &0.052 &0.32 \\
J1829+4844  &1828+487  &3C\,380   &0.692  &Q &$<$19.3&1  &... &...  &6.314 &19.49\\
J1944+5448  &1943+546  &OV\,573   &0.263  &G &20.7   &1  &315 &0.86 &0.149 &1.95 \\
J1945+7055  &1946+708  &...       &0.101  &G &21.5   &8  &357 &5.00 &0.063 &1.01 \\
J2022+6136  &2021+614  &OW\,637   &0.227  &Q &$<$19.6&1  &... &...  &0.031 &1.94 \\
J2052+3635  &2050+364  &...       &0.354  &G &20.9   &1  &16  &16.11&0.242 &4.51 \\
J2137$-$2042&2135$-$209&OX$-$258  &0.635  &G &$<$20.0&1  &... &...  &0.914 &4.86 \\
J2151+0552  &2149+056  &OX\,082   &0.74   &G &$<$21.2&2  &... &...  &0.016 &0.5  \\
J2250+1419  &2247+140  &4C\,14.82 &0.237  &Q &$<$20.1&1  &... &...  &0.630 &2.00 \\
J2344+8226  &2342+821  &...       &0.735  &Q &$<$19.8&1  &... &...  &1.006 &5.02 \\
J2355+4950  &2352+495  &OZ\,488   &0.2379 &G &20.5&1&82&1.72&0.199&2.24\vspace{0.2cm} \\
\hline\vspace*{-0.2cm}\\
\end{tabular}
\begin{tabular}{l}
\hi\ absorption reference: 1) Vermeulen et al.\ (\cite{vermeulen03}),
2) Carilli et al.\ (\cite{carilli98}), 3) van Gorkom et al.\
(\cite{vangorkom89}),\\4) Hagiwara pers.\ comm.: tentative detection
not yet confirmed, 5) Mirabel (\cite{mirabel89a}), 6)
Carilli et al.\ (\cite{carilli92}),\\ 7) Peck et al.\ (\cite{peck00})
and 8) estimated from the total integrated VLA spectrum, Peck et al.\ (\cite{peck99})\\
\end{tabular}
\end{flushleft}
\label{tablesample}
\end{table*}

We are interested in studying the \hi\ gas content of the AGN host
galaxy; both the general interstellar matter (ISM) in addition to
circumnuclear gas disks if present. In Table \ref{tablesample} we have compiled a
list of sources searched for \hi\ absorption which belong to the
GPS/CSS class and are therefore likely compact with total intrinsic
sizes $<$~10 kpc.

In order to achieve a sample of small sources searched for \hi\
absorption which is as complete and unbiased as possible, we have
selected sources from the following major GPS/CSS lists: Spencer et
al.\ (\cite{spencer89}), the Fanti CSS sample (Fanti et al.\
\cite{fanti90}), O'Dea \& Baum (\cite{odea97}), the CSS classified
objects from Morganti et al.\ (\cite{morganti97}), de Vries et al.\
(\cite{devries97a}) and the COINS sample (Peck et al.\
\cite{peck00}). The selection criteria were: \\1) The redshift must be
$0.05<z<0.75$. This ensures a high possibility of the objects having
been searched for \hi\ absorption, given the frequency coverage of the
WSRT (which has a UHF system covering the \hi\ 21~cm line up to
$z=0.75$). 2) The declination should be larger than $-30^{\circ}$; the
majority of GPS/CSSs are found at redshift between 0.1 and 1,
restricting observations to the frequency flexible Northern hemisphere
telescope WSRT. 3) The sources should have intrinsic linear sizes
$<10$ kpc, in order to represent a true GPS/CSS object. We have
therefore excluded a few objects (J1110+4817 and J1148+5924) which are
known to have a clear one-sided core-jet structure and are therefore
likely foreshortened by projection (Peck et al.\ \cite{peck00}). After
the initial publication from which we selected the sources, 4 objects
(J0432+4138, J0702+3757, J1227+3635 and J1520+2016) were found to
actually have redshifts $>$1.0 and thus have been excluded from our
analysis.

This gives a sample of 49 genuine GPS/CSS objects, which are listed in
Table \ref{table1}. We find that usable \hi\ absorption data have been
taken for 41 of these, most of which can be found in Vermeulen et al.\
(\cite{vermeulen03}).

Table \ref{table1} lists some properties of the sources included in
our analysis: the source J2000 and B1950 name, a possible other name,
redshift and the \hi\ column density with the corresponding
reference. All column densities assume a spin temperature of 100~K and
a covering factor of unity. Quoted upper limits are given as a
3$\sigma$ limit, assuming a line width of 100~\kms. For the sources
where the absorption line is best fitted by multiple Gaussians, the
number given is the total integrated column density. The FWHM and peak
opacity are listed for the component with the largest column density
if multiple lines are present. The projected linear size was (if
possible given the source radio morphology) derived from measuring the
size between the two hotspots in VLBI continuum images. For J0556-0241
we have only an upper limits of the size, however it is still included
in the following analysis. $S_{\rm cont}$ is the observed flux density
at the redshifted \hi\ frequency.  A Friedmann cosmology with $H_{\rm
0}=$~75~\kms\ and $q_{\rm 0}=~$0.5 is used throughout this paper; all
numbers taken from the literature are, if necessary, recalculated to
this cosmological model.


\section{Results}\label{results_stattot}

\subsection{Detection rate}\label{results}

Including all 41 targets with usable \hi\ data listed in Table
\ref{tablesample}, the \hi\ absorption detection rate is 54\%. We show
in Table \ref{optfrac} that 59\% of the galaxies have an \hi\
detection, as opposed to only 42\% of the quasars. The latter number
is however more uncertain since there are fewer quasars (29\%) in our
sample than galaxies (71\%). Interestingly, we find an even stronger
dichotomy when we separate out the GPS from the CSS sources by using a
linear size division at 1~kpc. Table \ref{gpsfrac} shows that the
smaller GPS sources have a detection rate of 63\%  as
compared to 36\% in the larger CSS sources; again the latter number is
more uncertain due to the smaller number of objects. Below, we will
discuss at length how this difference could
reflect the fact that the more compact sources have a larger part of
their continuum emission covered by nuclear gas. This effect has
already been suspected on account of the high detection rate of \hi\
absorption in nearby GPS/CSS objects (Conway \cite{conway96}). First,
rather than simply looking at the detection rate in two bins, we show
that there may be a more continuous relationship between 
\hi\ column density and linear size.

\begin{center}
\begin{table}[t]
\caption[]{Relation to optical identification.
\label{optfrac}}
\begin{tabular}{lccc}
\hline
\hline\vspace{-0.2cm}\\
 & Detected & Nondetected & Total \vspace{0.2cm}\\
\hline\vspace{-0.2cm}\\
Galaxies & 17 & 12 & 29 \\
Quasars  &  5 & 7  & 12 \vspace{0.2cm}\\
Total    & 22 & 19 & 41 \vspace{0.2cm}\\
\hline
\end{tabular}
\end{table}
\end{center}

\begin{center}
\begin{table}[t]
\caption[]{Relation to radio identification.
\label{gpsfrac}}
\begin{tabular}{lccc}
\hline
\hline\vspace{-0.2cm}\\
 & Detected & Nondetected & Total \vspace{0.2cm}\\
\hline\vspace{-0.2cm}\\
GPS & 17 & 10 & 27 \\
CSS &  5 &  9 & 14 \vspace{0.2cm}\\
Total& 22& 19 & 41 \vspace{0.2cm}\\
\hline
\end{tabular}
\end{table}
\end{center}


\subsection{Relation between $N_{\rm HI}$ and linear size}\label{nhisize}
\begin{figure}[t] \centering{
\resizebox{\hsize}{!}{\includegraphics{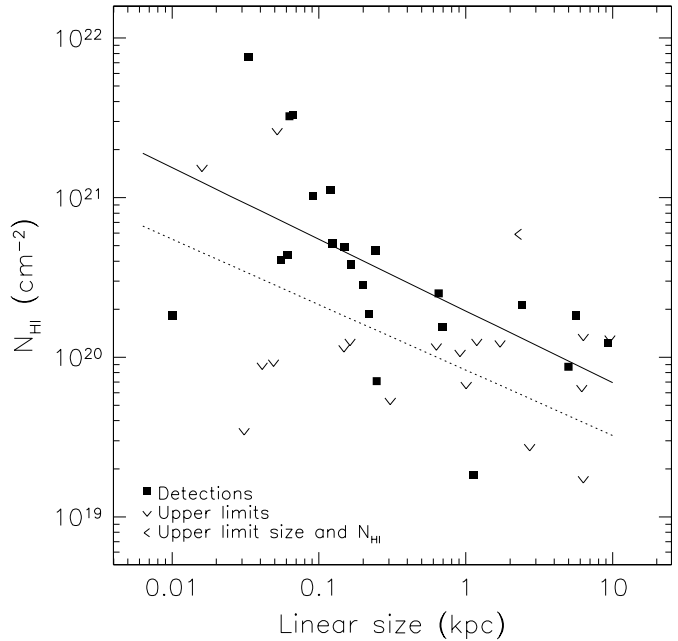}}
\caption{Absorbed \hi\ column density versus projected linear size.
There is an anti-correlation between the source size and the amount of
absorbing gas, confirmed by survival analysis. Best least square fit
taking into account the upper limits is shown with the dotted
line. This can be compared with a least square fit to the detections
only, plotted with the solid line.
\label{nhi_size}}} 
\end{figure}

There appears to be an anti-correlation between the source linear size
and the largest component of the \hi\ column density as plotted in
Fig.~\ref{nhi_size} (for the objects with multiple lines we plot the
component with the largest column density). A visual inspection
suggests that the column densities for sources $<$~0.5 kpc are larger
than for those $>$~0.5kpc. The data set contains upper limits both in
\hi\ column density, as well as a few points which have upper limits
on their size. We therefore investigated the possible correlation in
Fig.~\ref{nhi_size} using survival analysis, taking into account also
the upper limits. We used the software package ASURV Rev 1.2 (Lavalley
et al.\ \cite{lavalley92}), which implements the methods for bivariate
problems presented by Isobe et al.\ (\cite{isobe86}). A Kendall's Tau
test shows that there exists a correlation between the column density
and the linear size with a probability $>$~99\%.

In order to parameterise the correlation we fit the simplest model of
a power law to the data. The cluster of sources with column densities
about $10^{21}$ \cms\ suggests that other models are possible, perhaps
a two component model or a model with different power laws for small
and large sources. However, given the uncertainties of the present
dataset we cannot differentiate between these models. Linear
regression applied with ASURV finds the relationship $N_{\rm
HI}=7.2\times$\xp{19}$LS^{-0.43}$\cms; where $LS$ is the linear size
in kpc. This linear fit is plotted with a dotted line in
Fig.~\ref{nhi_size}. The solid line represents a least square linear
fit taking into account the detections only (parameterised by $N_{\rm
HI}=1.95\times$\xp{20}$LS^{-0.45}$\cms).

\begin{figure}[t] \centering{
\resizebox{\hsize}{!}{\includegraphics{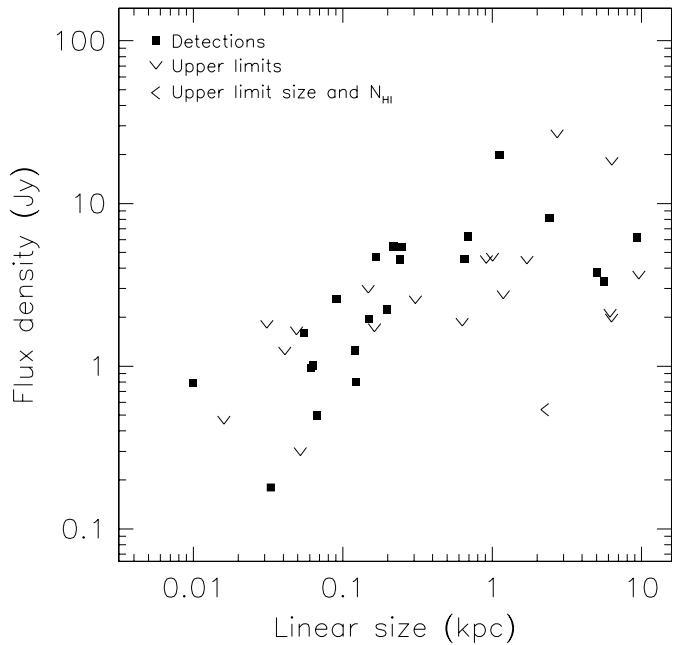}}
\caption{Observed continuum flux density at observed UHF frequency
versus projected linear size. This plot shows that the lack of large
sources with substantial absorption is not due to insufficient
sensitivity.}
\label{icont_size}} \end{figure}

While the majority of the upper limits do not show any dependence with
linear size, there are two of the most compact sources which have
higher upper limits. This raises the question whether the observed
anti-correlation between $N_{\rm HI}$ and $LS$ could be an
observational effect. Since the sensitivity of the \hi\ column density
depends on the strength of the background continuum, we plot in
Fig.~\ref{icont_size} the distribution of background continuum
strength as a function of linear size. Fig.~\ref{icont_size} shows
that we are more sensitive to small amounts of \hi\ absorption for the
larger sources. Clearly, there are no reasons why we should not have
been able to detect sources in the top right corner of
Fig.~\ref{nhi_size} if they existed, and so we are confident that the
top right hand corner on Fig.~\ref{nhi_size} is really empty. It is
however possible that there are additional absorbers amongst the
compact sources, with column densities similar to the larger sources
($\sim$~\xp{20} -- \xp{21}~\cms). This would reduce the amplitude of
the correlation (however the upper limit data and its possible
dependence on size has already been taken account of by ASURV when
making its confidence estimate).

\begin{figure}[t] \centering{
\resizebox{\hsize}{!}{\includegraphics{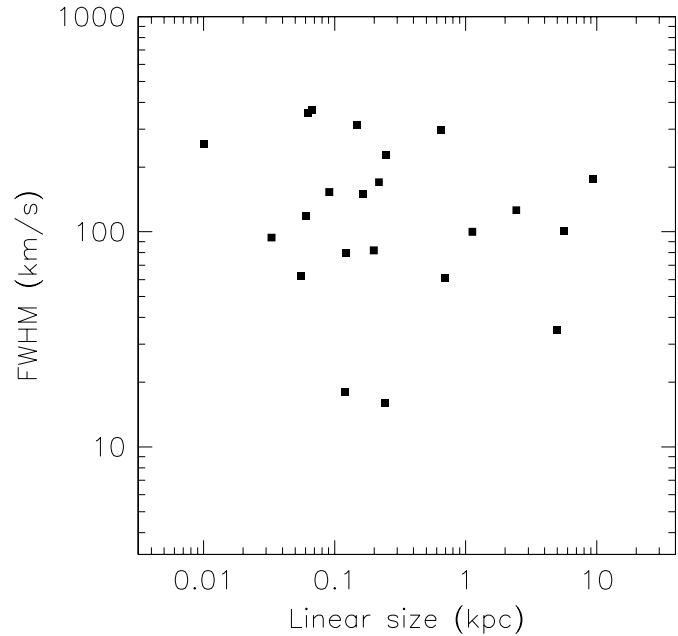}}
\caption{FWHM versus linear size. There is no correlation between the two
variables. \label{fwhm_size}} } \end{figure}

\begin{figure}[t] \centering{
\resizebox{\hsize}{!}{\includegraphics{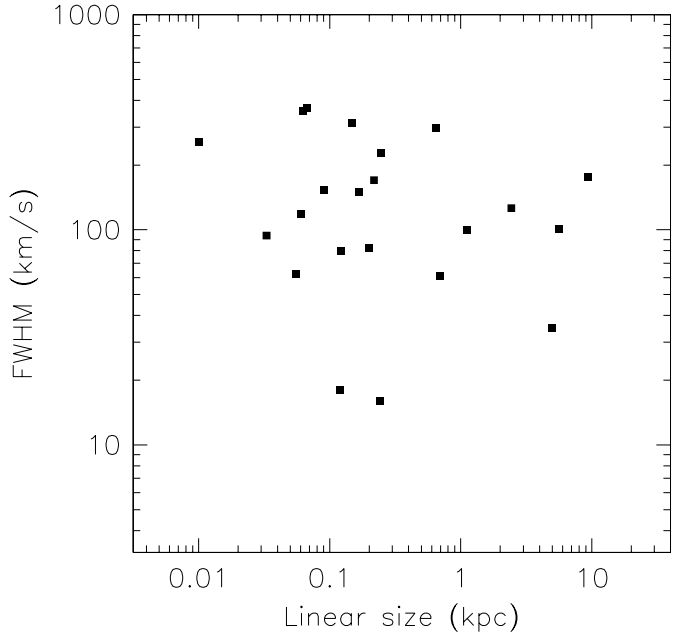}}
\caption{Peak optical depth versus linear size. A clear correlation is
found with a probability $>$~99\%, and the slope can explain most of the
correlation in Fig.\ \ref{nhi_size}. \label{tau_size}} } \end{figure}

An interesting question is whether the population of small sources
have larger column densities because the FWHM of the absorption line
is wider, or because the line is deeper. Fig.~\ref{fwhm_size} shows
the FWHM versus the projected linear size. There is no evidence that
there is any correlation between the two (tested by a Kendall's Tau
correlation test). Fig.~\ref{tau_size} shows the peak optical depth
versus linear size. A Kendall's Tau test shows that those two
variables are correlated with a probability $>$~99\%. The solid line
represents a least square linear fit to the data points. The slope is
$-$0.36, which implies that the slope in Fig.~\ref{nhi_size} mainly
depends on a difference in the observed opacity and is not due to
differences in the line width.

\subsection{A temporal change in the gas mass or a geometric effect?}
There are at least two plausible explanations for the observed
anti-correlation between source linear size and \hi\ column
density. The first possibility is that we are tracing a temporal
change in the total gas mass, and hence in the column density. In the
evolutionary models of GPS/CSSs the smallest sources are the
youngest, so we may postulate that the younger sources have used less
fuel during their lifetime, as compared to the older and larger
sources. To see if this is plausible, it is interesting to consider
the total gas mass and the rate of use for larger radio sources (e.g.\
FRIs and FRIIs). The radio loud sources are usually associated with
elliptical hosts (e.g.\ Martel et al.\ \cite{martel99}; Bahcall et
al.\ \cite{bahcall97}, Dunlop et al.\ \cite{dunlop03}), which perhaps is
surprising since it might be expected that more luminous sources would
reside in more gas rich systems.  However, the amount of gas needed to
sustain an AGN during its lifetime is not necessarily large. Consider
for instance a typical FRI radio galaxy with a bolometric luminosity
of the order of \xp{45}~\ergs. Given that these large sources have
expected lifetimes of $\sim$~\xp{7}~--~\xp{8}\,yrs (Parma et al.\
\cite{parma99}), and assuming an energy conversion efficiency of 10\%,
we would expect a fuel usage of less than \xp{7}\,\msun. Roughly 15\% of
the nearby normal ellipticals searched for \hi\ emission display \hi\
masses $>$~\xp{8}\,\msun\ (Huchtmeier \cite{huchtmeier94}). Amounts
smaller than \xp{7}~--~\xp{8}\,\msun\ are below present detection
limits, and are thus hard to detect in emission. Specifically in
ellipticals which harbour powerful AGN molecular gas masses of around
\xp{8}~--~\xp{9}\,\msun\ have been found in perhaps 40\% of the
sources searched during CO emission observations (Leon et al.\
\cite{leon01}), with detection limits of the order of \xp{8}\,\msun\
(Lim et al.\ \cite{lim00}). We therefore would expect total gas masses
of GPS/CSSs to be similar to that of classical FRI/FRII sources, given
the above mentioned expected fuel usage of $\la$~\xp{7}\,\msun\ and
that the typical total gas mass of FRI/FRIIs is
$\ga$~\xp{8}\,\msun. Hence, we reject the possibility of the
anti-correlation being due to a temporal change in the total gas mass, since
there is no evidence that the radio source would use up a significant
fraction of its total gas mass during its first \xp{4} years.

\begin{figure}[t] \centering{
\vspace*{0.2cm}
\rotatebox{-90}{\resizebox{\hsize}{!}{\includegraphics{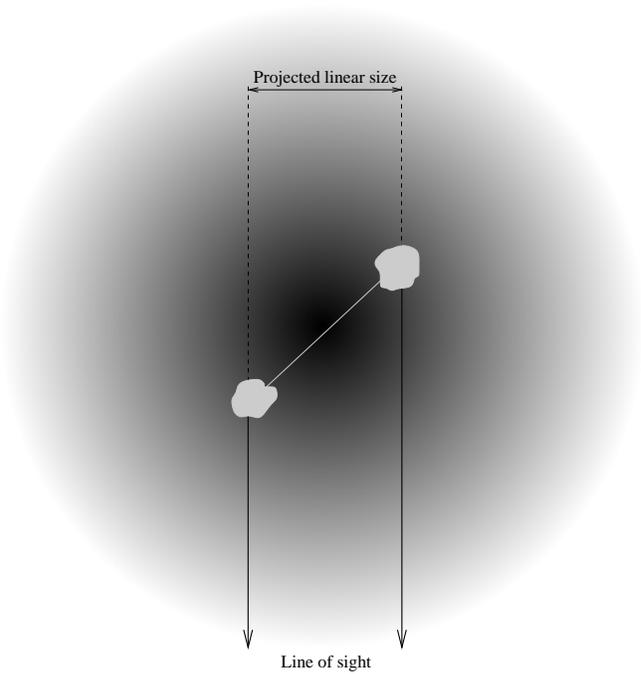}}}
\caption{Cartoon of the GPS/CSS geometry used for the modelling of a
spherical density distribution. The radio source is
confined within the host galaxy, for which we assume a maximum radius
of 50kpc, i.e.\ the ISM diameter is much larger than the largest
projected linear size of 10kpc. \label{csomodel1}}}
\end{figure}

Instead we prefer a second explanation, in which the observed
anti-correlation is a geometric effect of probing densities which
decrease with radial distance from the centre. Here we assume that the
sources (which all are $<10$ kpc) are fully embedded within the host
galaxy, and that the absorbing medium has basically the same
distribution in all sources (either a spherical or a disk
distribution; for a cartoon see Fig.\ \ref{csomodel1} and Fig.\
\ref{csomodel2}). If the radio source is centrally located in such a
gas distribution, the lobes of the smaller sources must be embedded in
(or in the case of a disk, behind) denser gas. Since most of those
objects are lobe-dominated we thus expect to probe larger foreground
column densities in the most compact sources. Constraining this
density profile is useful for constraining both the feeding mechanism
and evolution model of GPS/CSSs, and with this radial density model in
mind we will in the following section investigate the \hi\ absorption
characteristics.

\section{Distribution of the absorbing gas} 
\label{discuss_stat}

For some time it has been suspected that GPS/CSS sources have a high
\hi\ absorption detection rate (Conway \cite{conway96}). This seems to
be confirmed in our analysis of the available data (see
Section~\ref{results_stattot}). Most of those detections are so far
done using instruments of modest resolution, which means we cannot
easily determine where the \hi\ absorbing gas is located with respect
to the continuum. In the following discussion we will first consider
the case of the absorbing gas in our sample being in a spherical
distribution, then we discuss the possibility of the \hi\ absorption
instead arising in a disk.

\subsection{Spherical distribution} \label{spherical}

The strong radio emission in the GPS/CSS sources is most often
concentrated in two lobes, while the core emission (if present) is
weak. A cartoon of a GPS/CSS source geometry is shown in Fig.\
\ref{csomodel1}. The two lobes do not extend beyond the optical host
galaxy, hence the probability of tracing host galaxy ISM should be
high, especially if the central environment is enriched due to a
merger event. Assuming a spherical, radially declining density
distribution of the ISM gas, the larger sources will probe lower
density gas. In order to see if such a density distribution could
explain the anti-correlation in our data, we calculate the expected
integrated absorption $N_{\rm HI}=\int n(r)dl$ towards the two lobes,
where $dl$ corresponds to the path of absorption along the line of
sight. In this estimate we use a source with an inclination of
45$^\circ$, however, for this symmetric model the resulting \hi\
column density is only weakly dependent on the inclination.

\begin{figure}[t] \centering{
\resizebox{\hsize}{!}{\includegraphics{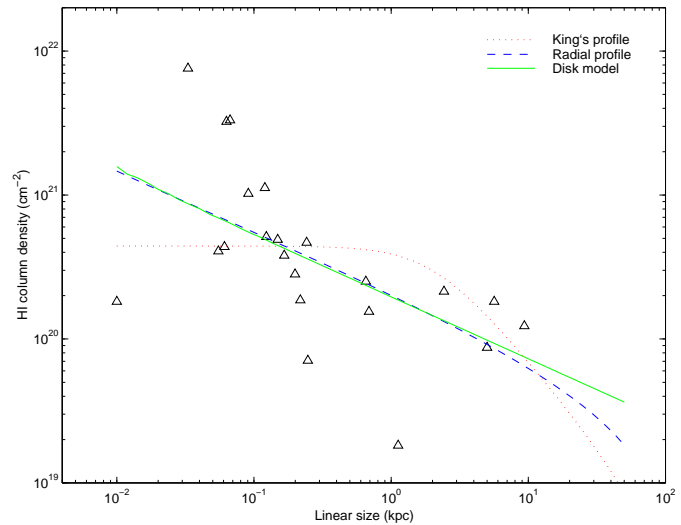}}
\caption{Results of calculating the expected \hi\ column density given
different density profiles. The triangles represents the real data. The King
density profile, plotted with a dotted line, cannot reproduce the same
\hi\ column density distribution for any value of $\beta$. Instead,
using a power law drop of the density provides a more similar
distribution, with the closest fit for $\beta$=~1.40 and $n_{\rm
0}=$~9\xxp{-3} \cmc\ (dashed line). Any kind of disk model can fit
the data reasonably well, here a disk opening angle of 20$^\circ$ is
plotted (solid line). \label{nmodels}}} \end{figure}

A first approach is to assume that the density follows a King profile:

\begin{equation}
n=n_{\rm 0} (1+\frac{r^{\rm 2}}{r_{\rm c}^{\rm 2}})^{\rm -\beta}
\label{kingeq}
\end{equation}

where $r_{\rm c}$ is the core radius of the gas distribution and
$n_{\rm 0}$ is is the density at 1 kpc. From X-ray data, which traces
the ionised gas component, the core radius for ellipticals is usually
derived to be $\sim$~1~kpc (Trinchieri et al.\ \cite{trinchieri86};
Forman et al.\ \cite{forman85}). We consider the simplest case, that
the atomic gas component is a constant fraction of the ionised gas and
follows a similar profile. The two variable parameters to be
constrained are $n_{\rm 0}$ and $\beta$, and we test the goodness of
the fit by calculating the sum of squares for the observed and
calculated \hi\ column densities. The best fit to the data is then
achieved for $n_{\rm 0}=$~0.05~\cmc\ and $\beta=$~1.06. The resulting
absorbed \hi\ column density as a function of source size is plotted
with the dotted line in Fig.\ \ref{nmodels} (calculated for a core
radius of 1~kpc and host ISM outer radius of 50~kpc). In this figure
the observed values in our sample are plotted with triangles. The King
model has a more or less uniform density in its centre, thus for any
value of $\beta$ such a density model is unable to reproduce the slope
in the data at small linear sizes. The only possibility to reproduce
the observed data is to reduce the core radius by several orders of
magnitude and, in addition, to increase the core density above
100~\cmc. However, then the density model does not follow a typical
King ISM profile any longer.

\begin{figure}[t] \centering{
\vspace*{0.5cm}
\rotatebox{-90}{\resizebox{\hsize}{!}{\includegraphics{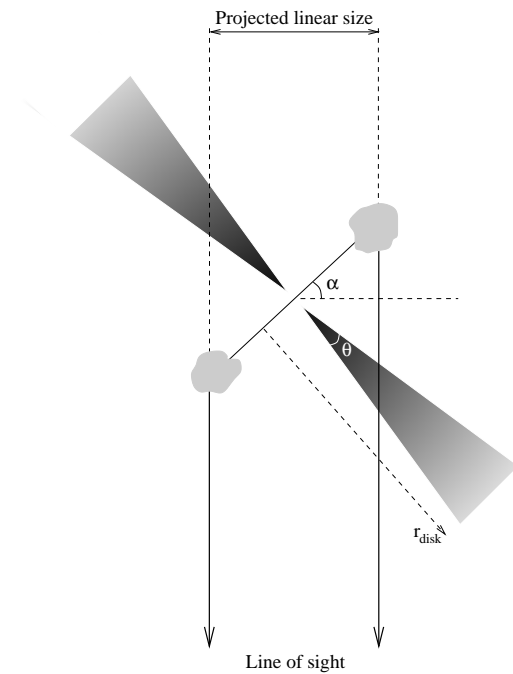}}}
\caption{Cartoon of the GPS/CSS geometry used for the disk modelling (not
to scale). The disk model is dependent on the disk opening angle
$\theta$ and the viewing angle $\alpha$, see Sect.\
\ref{disk}. \label{csomodel2}}}
\end{figure}

A King model of the density appears inconsistent with not only our
data, but also with what is generally assumed for the medium in which
the GPS sources expand. More often a simple, steep power law is used,
and such power laws successfully explain the detection rates of
sources of different sizes (e.g.\ Fanti et al.\ \cite{fanti95};
Readhead et al.\ \cite{readhead96}; O'Dea \& Baum
\cite{odea97}). Therefore we instead assume a power law density
profile 

\begin{equation}
n=n_{\rm 0}(\frac{r}{r_{\rm 0}})^{\rm -\beta}
\label{radialeq}
\end{equation} 

where $n_{\rm 0}$ is the number density (in \cmc) at $r_{\rm 0}=$~1
kpc. We again construct a sum of squares surface, which gives a best
fit to the data for $\beta\simeq$~1.40 and $n_{\rm 0}=$~0.009~\cmc
(this best fit is plotted in Fig.\ \ref{nmodels}). Assuming a volume
filling factor of unity, the integrated \hi\ mass within a radius of
10~kpc can then be estimated to be 7\xxp{7}~\msun. We note that if the
filling factor is less than 1, the estimated mass will decrease, and
our estimated \hi\ mass is thus an upper limit. Since the host
galaxies of GPS/CSSs may preferentially be found in
merging/interacting systems (de Vries et al.\ \cite{devries00}; O'Dea
et al.\ \cite{odea96}), it is interesting to compare the mass of \hi\
with what has been found in the more nearby population of the luminous
infra-red galaxies (ULIRGs). Those sources are also associated with
interactions (although between gas rich galaxies), and more or less
all of them are detected in \hi\ emission with masses in the range of
5\xxp{8} to 3\xxp{10}\,\msun, with typical values of the order of
\xp{9}\,\msun\ (Mirabel \& Sanders \cite{mirabel88}). Even though our
objects may also be merging systems, a spherical distribution
indicates that the GPS/CSSs are less gas rich than the ULIRGs.

Given that the host galaxies of radio loud sources are ellipticals, it
is also interesting to compare our derived gas mass with what is found
in nearby non-AGN ellipticals. Around 15\% of those normal ellipticals
are detected in \hi\ emission with masses $M_{\rm
HI}\ga$~\xp{9}\,\msun\ (Huchtmeier et al.\ \cite{huchtmeier95}; Knapp
et al.\ \cite{knapp85}). Typical upper limits of the remaining 85\%
are of the order of \xp{8}~--~\xp{9}\,\msun. Given the sensitivity of
the \hi\ emission experiments, the \hi\ mass we calculate assuming a
spherical power law density profile is consistent with our GPS/CSS
sources being hosted by normal ellipticals.

\subsection{Disk distribution}\label{disk}

Instead of spherically distributed gas, another plausible distribution
is a disk in a plane roughly perpendicular to the radio source. For
most source orientations this means that only the far radio lobe is
occulted.  It has been shown from VLA and ATCA \hi\ imaging that
early-type E and S0 galaxies often have their \hi\ in disk systems
(e.g.\ Sadler et al.\ \cite{sadler00}; Oosterloo et al.\
\cite{oosterloo99}).  Many large scale radio sources show \hi\
absorption that is consistent with circumnuclear gas, extended on
scales of a few pc in radio loud FRIs up to scales of a few 100 pc in
low luminosity Seyferts (e.g.\ van Langevelde et al.\
\cite{vanlangevelde00}; Taylor \cite{taylor96}; Gallimore et al.\
\cite{gallimore99}; Vermeulen et al.\ \cite{vermeulen02}).  HST
imaging has revealed nuclear disks of gas and dust in radio loud AGN
(Verdoes Kleijn et al.\ \cite{verdoes99}; de Koff et al.\
\cite{dekoff00}). In particular recent HST observations of a few
GPS/CSS sources have shown them to contain central disks of gas and
dust (4C\,31.04, 1946+708 and 1146+596; Perlman et al.\
\cite{perlman01}).  Other high resolution absorption studies of nearby
GPSs show that the \hi\ data are consistent with a rotating ring
(e.g.\ in 1946+708; Peck et al.\ \cite{peck99} and 4C\,31.04; Conway
\cite{conway96}). Here we consider the case of the absorbing gas in a
disk oriented perpendicular to the radio axis.

In this case absorption will only be seen against the far lobe. For a
cartoon image of the simple disk see Fig.\ \ref{csomodel2}. The
maximum radius of the disk, $r_{\rm disk}$, is assumed to be at least
10 kpc and thus for sources of projected linear sizes up to 10 kpc the
far lobe is always covered by the disk.  When calculating the \hi\
column densities, we assume the disk has a radial density profile
$n=n_0 (r/r_{\rm 0})^{-\beta}$, where $n_{\rm 0}$ is the density at
$r_{\rm 0}=1$ kpc. The calculated column densities will be a function
of both viewing angle as well as the disk opening angle. In our
modelling we limit ourselves to a viewing angle of 45$^{\circ}$, and
uses disk opening angles of 10\degr, 20\degr\ and 30\degr. We again
test the goodness of fit using the sum of squares method. Given the
scatter in the data all such opening angles give similar fits, and in
Fig.\ \ref{nmodels} we plot the best fit for a disk full opening angle
of 20\degr, yielding $n_0=0.11$ and $\beta=1.45$.

A 10kpc radius disk will, assuming there is an inner radius of the
disk of 1pc, yield an atomic mass of the order of \xp{8}\,\msun. This
is comparable to the results of the spherical distribution (Sect.\
\ref{spherical}). Following the discussion in Sect.\ \ref{spherical},
we find that also for a disk distribution the GPS/CSSs appear to be
less gas rich than the ULIRGs.


\section{Comparison with GPS/CSS models}\label{models}

Currently the most favoured explanation to the small sizes of most GPS
and CSS sources is that they are young, as estimated both from their
measured expansion speed as well as from synchrotron ageing (e.g.\
Owsianik \& Conway \cite{owsianik98a}; Murgia et al.\
\cite{murgia99}). In this 'youth' scenario the dense gas interacts
with the radio plasma giving a large radio luminosity which then
explains the relatively larger number of sources with small size.  In
order to explain the number density of sources of a given size, the
radio jet is assumed to propagate in a medium with decreasing density,
approximated with a power law profile $\propto r^{\rm -\beta}$ with a
slope $\beta=$~1.5~--~2 (Fanti et al.\ \cite{fanti95}; Begelman
\cite{begelman96}, Readhead et al.\ \cite{readhead96}). Given our
uncertainties, it is interesting that this exponent is comparable with
the $\rm \beta\simeq$~1.4 we get from our \hi\ absorption. 

Alternatively the GPS/CSSs have been suggested to be old 'frustrated'
sources prevented from growing by a very dense medium (e.g.\ van
Breugel et al.\ \cite{vanbreugel84}; Fanti et al.\ \cite{fanti90}). In
Sect.\ \ref{intro_stat} we mentioned several observations showing the
presence of gas in GPS/CSSs. Other evidence for dense environments
comes from effects thought to derive from interactions with the
surrounding medium, several CSS objects for example display distorted
radio morphologies (e.g.\ van Breugel et al.\ \cite{vanbreugel84}) and
alignment effects between the radio axis and the optical emission
(e.g.\ de Vries et al.\ \cite{devries97b}). Obviously, there is
compelling evidence for dense gas in GPS/CSS sources; the question is
whether the central density is large enough to give frustration.

While the frustration model is now less popular it is still possible
that a minority of the GPS/CSS sources are created in this way. Here
we discuss our \hi\ results in context of those two models.

\subsection{The frustration scenario}
\begin{table*}[t]
\begin{center}
\caption[]{Density of the medium where the lobes expand, as expected
from lobe advance speed ($n_{\rm ext}$), compared to \hi\ density
estimates from a spherical density model ($ n_{\rm S,100K}$) and a
disk model ($n_{\rm D,1000K}$). Estimates are done using $h=0.75$. The
first part of the table lists sources which are detected in \hi\
absorption, while the second part lists sources that are not detected
in \hi\ absorption.
\label{expdens}}
\begin{tabular}{llllllllll}
\hline
\hline\vspace{-0.2cm}\\
&Source & radius & $v_{\rm a}$ & $p_{\rm i}$ &  ref$^{\rm
1}$ & $ n_{\rm ext}$ & $C_{\rm HI}$ & $ n_{\rm S,100K}$ & $ n_{\rm D,1000K}$ \\
&& kpc &$ h^{\rm -1}\,c$ & $ h^{\rm 4/7}$dy\,\cms& \cmc& & \cmc& \cmc \vspace{0.2cm}\\
\hline\vspace*{-0.2cm}\\
Sources with            &J0111+3906& 0.017& 0.098 & 1.1\xxp{-4} & 1   & 3.6 & 9.1 & 24.6 & 368.4 \\
detected \hi\ absorption&J1407+2827& 0.005& 0.050 & 2.5\xxp{-4} & 2,3 &31.8 & 0.1 & 1.5  &  23.9 \\
                        &J1944+5448& 0.087& 0.210 & 2.5\xxp{-4} & 4,5 & 1.8 & 1.1 & 0.3  &  4.2 \\ 
                        &J2355+4950& 0.078& 0.101 & 9.7\xxp{-5} & 6,7 & 3.0 & 0.7 & 0.2  &  3.1 \vspace*{0.2cm} \\
\hline\vspace*{-0.2cm} \\
Sources without         &J0713+4349& 0.058& 0.126 & 1.9\xxp{-4} & 8,7 & 3.8 & ... & 0.5  & 6.8 \\
detected \hi\ absorption&J1035+5628& 0.082& 0.310 & 2.9\xxp{-5} & 9   & 0.1 & ... & 0.3  & 4.1 \\
                        &J2022+6136& 0.011& 0.060 & 6.4\xxp{-4} & 10  &56.5 & ... & 5.0  & 76.1 \vspace*{0.2cm}  \\
\hline\vspace*{-0.1cm}\\
\end{tabular}
\begin{tabular}{l}

$^{\rm 1}$References for lobe advance speed and the lobe pressure: 1)
Owsianik et al.\ \cite{owsianik98b}, 2) Polatidis \& Conway
\cite{polatidis03},\\ 3) Stanghellini et al.\ \cite{stanghellini97},
4) Polatidis et al.\ \cite{polatidis99}, 5) A.\ Polatidis, pers.\
comm., 6) Owsianik et al.\ \cite{owsianik99},\\ 7) Conway et al.\
\cite{conway92}, 8) Owsianik \& Conway \cite{owsianik98a}, 9) Taylor
et al.\ \cite{taylor00} and 10) Tschager et al.\ \cite{tschager00}
\end{tabular}
\end{center}

\end{table*}

In order to confine a median luminosity CSS, an average number density
in the range 1~--~10 \cmc\ at $r=$~10 kpc is needed (de Young
\cite{deyoung93}). This implies, for a uniform density, gas masses of
the order of \xp{10}\,\msun. In our power law density model we derive
atomic gas masses of the order of \xp{8}\,\msun, two orders of magnitude
less. Comparison of molecular data with \hi\ data for elliptical
galaxies have indicated a $M_{\rm H2}/M_{\rm HI}$ ratio of the order
of 0.4 (Wiklind et al.\ \cite{wiklind95}). If such a number is
applicable also to the GPS/CSS sources, it implies total gas masses
around a few times \xp{8}\,\msun. This result may change if the ratio
$M_{\rm H2}/M_{\rm HI}$ is very different. For the gas rich ULIRGs
however, the total molecular gas mass derived from CO emission, and
the atomic gas mass derived from \hi\ emission are of the same order
(\xp{9}\,\msun; Downes \& Solomon \cite{downes98}; Mirabel \& Sanders
\cite{mirabel88}), and will thus change the resulting total gas mass
by a small factor.  Given those estimates, the \hi\ absorption data
suggests that there is not enough gas to frustrate the radio sources.


\subsection{The youth model}
The young ages estimated from the hotspot advance speeds are strong
arguments in favour of the youth model. Using ram pressure arguments,
the lobe advance speed $v_{\rm a}$ is related to the external mass
density $\rho_{\rm ext}$ and the hotspot pressure $ p_{\rm i}$ by

\begin{equation}p_{\rm i}=\rho_{\rm ext}v_{\rm a}^{\rm 2}=n_{\rm ext}m_{\rm p}v_{\rm
a}^{\rm 2}
\label{hotspotp}
\end{equation}

Thus, for sources with measured expansion velocities, we can estimate
the density $n_{\rm ext}$ at the hotspot location. We can further
compare $n_{\rm ext}$ with the density expected using the power law
density distribution models discussed in Sect.\ \ref{discuss_stat}.

Published in the literature we find a selection of 7 sources with
measured advance speeds and quoted (or sufficient information to
calculate) internal hotspot pressures. For those sources, we list in
Table \ref{expdens} the distance between the core and hotspot, the
advance speed, hotspot pressure and the corresponding references. We
use Eq.\ (\ref{hotspotp}) to derive external densities $n_{\rm ext}$
(Table \ref{expdens}) at the given hotspot radius.

In order to derive densities using the disk and spherical power law
density profile (Eq.\ (\ref{radialeq})), we first note that all sources
listed in Table \ref{expdens} have very small linear
sizes. Especially, for sources $<0.1$ kpc there are large offsets
between the measured \hi\ column density and the \hi\ column density
as predicted by our simple models (Fig.\ \ref{nmodels}). For the
sources with measured \hi\ column density, we therefore apply a
correction factor ($C_{\rm HI}=N_{\rm HI, measured}/N_{\rm HI,
predicted}$), in order to better estimate the density in the source in
question. We also note that the estimated density will depend on the
spin temperature; the values for the column density used in Sect.\
\ref{results_stattot} assumes a $T_{\rm sp}=100$K. However, close to
an X-ray irradiating AGN the spin temperature is likely to be above
100K, and probably closer to 8000 K if the gas is purely atomic
(Maloney et al.\ \cite{maloney96}).

For the spherical density profile, we use the power law profile (Eq.\
(\ref{radialeq})) and express the density $n_{\rm S}$ as a function of
radius and spin temperature:

\begin{equation}
n_{\rm S}= C_{\rm HI}n_{\rm HI}(\frac{T_{\rm sp}}{100\rm K})=C_{\rm HI}n_{\rm 0}r_{\rm
 kpc}^{-\beta}(\frac{T_{\rm sp}}{100\rm K})
\end{equation}

Applying the best fitting values of $n_0$ and $\beta$ from
Sect.~\ref{spherical}, and using a spin temperature of 100K we
achieve:

\begin{equation}
n_{\rm S,100K}\simeq C_{\rm HI}$~0.009$r_{\rm kpc}^{\rm -1.4}
\label{ns}
\end{equation}

At a given radius Eq.\ (\ref{ns}) gives the \hi\ density, and is thus a
lower limit to the total density. Using the radius of the sources in
Table \ref{expdens}, we calculate the corresponding values of the
density $n_{\rm S,100K}$. For this model we have chosen to use $T_{\rm
sp}=100$K, since that gives the closest match to the densities $n_{\rm
ext}$ estimated from the ram-pressure argument. If the spin
temperature is 8000 K, the estimated $n_{\rm S}$ will be 80 times
higher.

If instead the absorbing gas is in a disk distribution, we can
estimate the density at the hotspots (which must be in a medium
external to the disk, see Fig.\ \ref{csomodel2}) by assuming that the
disk internal pressure will be similar to the pressure outside the
disk. Since we do not have any information on the properties of such a
disk, we will first consider the simplest case, assuming the disk to
be completely atomic and to have an uniform kinetic temperature
$T_{\rm disk}=T_{\rm sp}$. The external medium is assumed to be
ionised with an electron temperature $T_{\rm ext}=$~\xp{4} K. Applying
the ideal gas law the external density $n_{\rm D}$ can be written as:

\begin{equation}
n_{\rm D}= C_{\rm HI}n_{\rm HI}(\frac{T_{\rm sp}}{100\rm K})(\frac{T_{\rm
sp}}{T_{\rm ext}})=C_{\rm HI}n_{\rm 0}r_{\rm kpc}^{\rm
-\beta}(\frac{T_{\rm sp}}{100\rm K})(\frac{T_{\rm sp}}{T_{\rm ext}})
\label{eq3}
\end{equation}

For a disk opening angle of 20\degr, we derived a value of $n_{\rm
0}=$~0.11 \cmc\ and $\beta=$~1.45 (Sect.~\ref{disk}), and assuming a
spin temperature of 1000K we rewrite Eq.\ (\ref{eq3}) into

\begin{equation}
n_{\rm D,1000K}\simeq ~0.11~ C_{\rm HI}r_{\rm kpc}^{\rm -1.45}
\end{equation}

Again, this is a lower limit to the total gas density present. In
order to get densities which best agree with those estimated via the
ram pressure argument, we here use a $T_{\rm sp}=1000$K. The resulting
values of $n_{\rm D,1000K}$ are listed in Table \ref{expdens}.

Excluding J0111+3906, for the spherical power law model most $n_{\rm
ext}$ and $ n_{\rm S,100K}$ values are remarkably similar, and in most
cases only differ within an order of a magnitude; which is consistent
with the GPSs propagating through a purely atomic spherical gas
distribution. The offset can depend on a number of things. Firstly, it
is possible that the spin temperature has a very different
value. Secondly, and closely linked to the spin temperature, is the
expected molecular abundance. If the kinetic temperature is as low as
100K, the molecular abundance is expected to be high (Maloney et al.\
\cite{maloney96}), and thus our \hi\ density estimate is only a lower
limit to the total gas density. Thirdly, the value of $\beta$ and
$n_{\rm 0}$ could be different, given the sparse amount of data used
in the fitting (22~points). Within a 90\% confidence region the value
of $\beta$ may change as much as $\pm$0.25 and $n_{\rm 0}$ as much as
$\pm$4e-3, with a corresponding difference of $\sim$~1 in the number
density at a radius of 0.05 kpc. Finally the individual viewing angles
of the sources in Table \ref{expdens} may affect the assumed radius
where the hotspot is located.

The densities derived from the disk profile are equally similar,
differing with only a factor of a few from $ n_{\rm ext}$ (except for
J0111+3906). The offsets are not surprising, given the large number of
uncertainties in the calculation of $n_{\rm D,1000K}$. In addition to
the reasons already given for the spherical distribution, the ratio
$T_{\rm disk}/T_{\rm ext}$ could for instance be very different from
the assumed value.

We conclude that within the scope of the power law profile, we more or
less can reproduce the densities expected in the youth model for the
GPS sources with measured lobe velocities.


\section{Orientation effects on the observed HI\\ column
density: galaxies versus quasars}\label{orientation}

\begin{figure}[th!]
\centering{
\resizebox{\hsize}{!}{\includegraphics{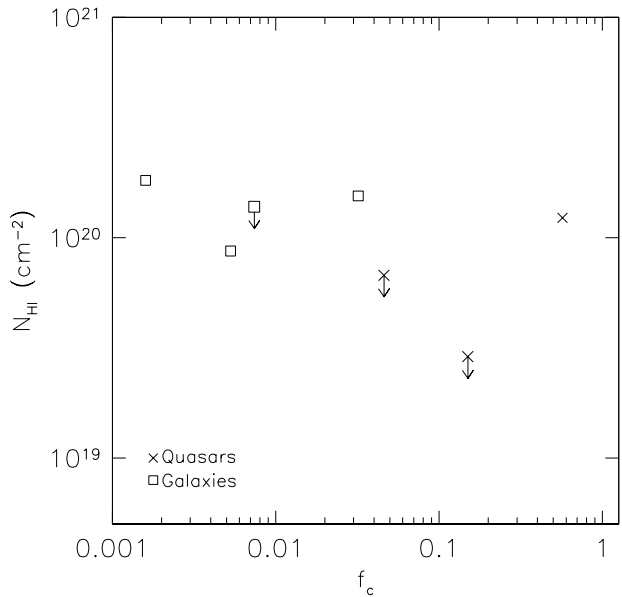}} \caption{\hi\
column density versus fraction of core emission $f_{\rm
c}$. Quasars are marked with crosses, and galaxies with squares. In
general the quasars are found in sources with larger core prominence,
consistent with smaller viewing angles.}
\label{fcfig}}
\end{figure}

In most cases the spatial resolution of present \hi\ data is not
enough to determine whether the absorption covers both lobes
(consistent with ISM gas) or only one lobe (indicating a disk
distribution). In the previous sections we showed that, given the
limited information we have about the density profiles, both a disk as
well as a spherical distribution are consistent with the observed
relation between \hi\ column density and linear size. One possibility
to distinguish between those models is to look at any possible
orientation effects. Assuming the unified scheme, quasars are supposed
to be seen at a smaller viewing angle $\alpha$ (more end-on) than the
radio galaxies. In fact, we have shown that there appears to be a
larger detection rate in galaxies than in quasars (Sect.\
\ref{results}).

For a given projected source size of 1 kpc, the amount of expected
\hi\ absorption as a function of viewing angle can be estimated. Given
that for a projected size of 1 kpc we have a \hi\ column density
sensitivity of around \xp{20}~\cms\ (Fig.~\ref{nhi_size}), the
spherical density profile predicts we would detect equal column
densities in galaxies (with say $\alpha\ga$~45\degr; Barthel
\cite{barthel89}) and quasars ($\alpha\la$~45\degr). In contrast for
the disk profile there will be a range of optimum viewing angles, for
which we are likely to detect \hi\ absorption. This exact range of
viewing angles will be extremely dependent on the disk model used.  In
Sect.\ \ref{fc} we discuss if the galaxies and quasars in our sample
are consistent with having different viewing angles, as expected from
unified schemes. If so, a spherically symmetric distribution could be
considered less likely, since objects classified as galaxies appear to have
higher \hi\ column densities than quasars.

\subsection{Core prominence}\label{fc}

Because of relativistic boosting the core strength is considered to be
a good indicator of orientation; the more dominant the core is, the
smaller the viewing angle.  For larger radio sources, this has been
investigated by a number of people (e.g.\ Orr \& Browne
\cite{orr82}). Since the GPS/CSS objects in general have weak or no
core emission, similar studies have been difficult. Stanghellini et
al.\ (\cite{stanghellini01}) reports on VLBA observations of 10 GPS
sources. The authors conclude that in the few cases where cores were
detected, the core of the GPS galaxies accounted for $\simeq 2$\% of
the total flux density, while in the GPS quasars the core
accounted for $\simeq 20$\%. Moreover, Saikia et al.\ (\cite{saikia95})
studied a sample of CSS sources with detected radio cores, and found
that the degree of core prominence were consistent with those for
larger radio sources and also consistent with the unified scheme. In
the study by Saikia et al.\ (\cite{saikia95}), there are 7 sources
that can also be found in our sample (3 quasars and 4 galaxies). For
the galaxies the average fraction of emission from the core is
$\overline{f}_{\rm c}=$~0.012, while the corresponding number for the
quasars is $\overline{f}_{\rm c}=$~0.26. In Fig.~\ref{fcfig} we plot
the \hi\ column density versus $f _{\rm c}$ for those objects, and we
mark quasars and galaxies with different symbols. Clearly the quasars
have larger core prominence. This implies the quasars have a smaller
viewing angle.

Only one of these quasars is detected in \hi\ absorption, while three
of the galaxies are \hi\ absorbers.  For the sources in common with
the sources of Saikia et al.\ (\cite{saikia95}), it appears as if the
upper limits of the \hi\ column density for the quasars are smaller
than the detections for the radio galaxies. It would be interesting to
carry out deeper observations for a larger sample of the CSS quasars
with upper limits; if deeper searches showed (for a majority of the
sources) lower column densities or upper limits for the quasars than
for the radio galaxy detections, this would strongly argue for a disk
origin for the \hi\ absorption.

\subsection{GPS quasars versus GPS galaxies}\label{qvsg}
It should be noted that it is debated whether the GPS quasars are
really intrinsically small sources like the GPS galaxies. Snellen
(\cite{snellen97}) proposes that GPS galaxies viewed end-on will have
a flat spectra and thus the GPS galaxies and quasars are different
types of objects. This has some support in a very different redshift
distribution for the GPS galaxies and quasars (Snellen
\cite{snellen97}; Stanghellini \cite{stanghellini98}), and the fact
that many GPS quasars appear to have core-jet morphologies when
observed at high angular resolution (Stanghellini
\cite{stanghellini01}). If this is the case, the low detection rate in
the GPS quasars reported here could be explained by them being large
scale radio sources, hence with a lower probability of the line of
sight going through the host galaxy.


\section{Implications for the spectral turn-over mechanism}\label{turnover}

The integrated radio spectrum turn-over frequency for GPS/CSS sources
is known to have a relationship $\nu_{\rm t}\propto LS^{-0.65}$ (O'Dea
\& Baum \cite{odea97}). Most often this is ascribed to synchrotron
selfabsorption (e.g.\ O'Dea \& Baum \cite{odea97}); however an
alternative possibility is the one of free-free absorption (e.g.\
Bicknell et al.\ \cite{bicknell97}). A few VLBI results indicate the
presence of a nuclear disk causing free-free absorption (Marr et al.\
\cite{marr01}; Peck et al.\ \cite{peck99}).

Since our \hi\ results show that the column density along the line of
sight depends on the source linear size (and thus turnover frequency),
it is likely that a similar relationship exists for an ionised gas
component. The free-free absorption turn-over frequency is related to
the electron density, $\nu _{\rm t} \propto (n_{\rm e}^2)^{1/2.1}$. If
we simply assume the gas fraction to be constant ($n_{\rm HI}/n_{\rm
e}=C$) and $N_{\rm HI}=n_{\rm e}L$ where L is the absorbing
pathlength, the observed relationship ($N_{\rm HI}\propto LS^{-0.43}$)
implies a free-free absorption caused turnover relationship $\nu_{\rm
t}\propto LS^{-0.41}$. This is much less steep than observed
($\nu_{\rm t}\propto LS^{-0.65}$), implying that free-free absorption
can not be the only turn-over mechanism present.

Conversely we could assume that the observed relationship between
turnover frequency and linear size is due to free free
absorption. Then, $\nu_{\rm t}\propto LS^{-0.65}$ implies that the
slope for the ionised gas component necessarily is larger than the
slope we observe for \hi. In this case perhaps there is a gradient of
the fraction of ionised gas decreasing with radius.


\section{Conclusions}
We have investigated the combined results of \hi\ absorption in
GPS/CSS sources. The most striking result is that the smaller GPS
sources ($<$~0.5 kpc) tend to have larger \hi\ column densities than
the larger CSS sources ($>$~0.5 kpc). This anti-correlation between
linear size and absorbed \hi\ column density is mostly an effect from
opacity differences rather than line width differences. It is possible
to use a simple power law profile of the density to explain the
observed \hi\ column densities, and both a spherically symmetric
medium as well as disk can be fitted to the data.

Our modelling cannot distinguish between the spherical and disk
density models, and most observations at present do not have enough
resolution to determine the location of the absorbing gas. However, to
date high resolution \hi\ data exist for a limited number of sources
that in all cases indicate \hi\ disks (Conway \cite{conway96}; Peck et
al.\ \cite{peck99}; Peck \& Taylor \cite{peck98}). This argues for a
disk model, and in addition we notice that most \hi\ absorbers are
found in sources classified as galaxies. By studying our sources in
their radio continuum properties we find that the galaxies and quasars
in our sample behave similarly to larger scale sources, where
orientation effects are known to take place. Therefore, it is easier
to argue for a disk model, where sources with very small viewing
angles in addition to sources observed edge-on would be less likely to
have \hi\ detectable in absorption against their lobes. The fact that
quasars have smaller column densities argues against a spherically
symmetric model for the \hi\ gas.

\begin{acknowledgements}
This research has made use of the NASA/IPAC Extragalactic Database
(NED) which is operated by the Jet Propulsion Laboratory, California
Institute of Technology, under contract with the National Aeronautics
and Space Administration.
\end{acknowledgements}


\begin{thebibliography}{}
\bibitem[1997]{bahcall97}
Bahcall, J. N., Kirhakos, S., Saxe, D. H., \& Schneider D.P. 1997,
ApJ, 479, 642

\bibitem[1989]{barthel89}
Barthel, P. D. 1989, ApJ, 336, 606

\bibitem[1996]{begelman96}
Begelman, M. C. 1996, in Cygnus A - a Study of a Radio Galaxy,
ed. C. L. Carilli, \& D. E. Harris, Cambridge University Press, 209

\bibitem[1997]{bicknell97}
Bicknell, G. V., Dopita M. A., \& O'Dea, C. P. 1997, ApJ, 485, 112

\bibitem[1992]{carilli92}
Carilli, C. L., Perlman, E. S., \& Stocke J. T. 1992, ApJ, 400, L13

\bibitem[1998]{carilli98}
Carilli, C. L., Menten, K. M., Reid, M. J., Rupen, M. P., \& Yun,
M. S., 1998, ApJ, 494, 175

\bibitem[1992]{conway92}
Conway, J. E., Pearson, T. J., Readhead, A. C. S., et al. 1992, ApJ,
396, 62

\bibitem[1996]{conway96}
Conway, J. E. 1996, in The Second Workshop on Gigahertz Peaked
Spectrum and Compact Steep Spectrum Radio Sources, ed. I. Snellen,
R.T. Schilizzi, H. A. J. R\"ottgering, \& M. N. Bremer, Publ JIVE,
Leiden, 198

\bibitem[2000]{dekoff00}
de Koff, S., Best, P., Baum, S. A., et al. 2000, ApJS, 129, 33

\bibitem[1997a]{devries97a}
de Vries, W. H., Barthel, P. D., \& O'Dea, C. P. 1997, A\&A, 321, 105

\bibitem[1997b]{devries97b}
de Vries, W. H., O'Dea C.P., Baum, S. A., et al. 1997, ApJS, 110, 191

\bibitem[2000]{devries00}
de Vries, W. H., O'Dea, C. P., Barthel P.D., et al. 2000, ApJ, 120, 2300

\bibitem[1993]{deyoung93}
de Young, D. S., 1993, ApJ, 405, L13

\bibitem[1998]{downes98}
Downes, D., \& Solomon, P. M. 1998, ApJ, 507, 615

\bibitem[2003]{dunlop03}
Dunlop, J. S., McLure, R. J., Kukula, M. J., Baum, S. A., O'Dea, C. P.,
\& Hughes, D. H. 2003, MNRAS in press, see astro-ph/0108397

\bibitem[1994]{elvis94}
Elvis, M., Fiore, F., Wilkes, B., McDowell, J., \& Bechtold, J. 1994,
ApJ, 422, 60

\bibitem[1999]{evans99}
Evans, A. S., Kim, D. C., Mazzarella, J. M., Scoville, N. Z., \& Sanders,
D. B. 1999, ApJL, 521, L107

\bibitem[1990]{fanti90}
Fanti, R., Fanti C., Schilizzi, R. T., et al. 1990, A\&A, 231, 333

\bibitem[1995]{fanti95}
Fanti, C., Fanti, R., Dallacasa, D., et al. 1995, A\&A, 302, 317

\bibitem[1985]{forman85}
Forman, W., Jones, C., \& Tucker, W. 1985, ApJ, 293, 102

\bibitem[1999]{gallimore99} 
Gallimore, J. F., Baum, S. A., O'Dea, C. P., Pedlar, A., \& Brinks, E. 1999,
ApJ, 524, 684

\bibitem[1994]{gelderman94}
Gelderman, R., \& Whittle, M. 1994, ApJS, 91, 491

\bibitem[1994]{huchtmeier94}
Huchtmeier, W. K. 1994, A\&A, 286, 389

\bibitem[1995]{huchtmeier95}
Huchtmeier, W. K., Sage, L. J., \& Henkel, C., 1995, A\&A, 300, 675

\bibitem[1986]{isobe86}
Isobe, T., Feigelson, E. D., \& Nelson, P. I., 1986, ApJ, 306, 490

\bibitem[2000]{kameno00} 
Kameno, S., Horiuchi S., Shen, Z.-Q., et al. 2000, PASJ, 52, 209

\bibitem[1987]{kato87}
Kato, T., Tabara, H., Inoue, M., \& Aizu, K. 1987, Nature, 329, 223

\bibitem[1985]{knapp85}
Knapp, G. R., Stark, A. A., \& Wilson, R. W. 1985, AJ, 90, 254

\bibitem[1992]{lavalley92}
Lavalley, M., Isobe, T., \& Feigelson, E. 1992, BAAS, 24, 839

\bibitem[2001]{leon01}
Leon, S., Lim, J., Combes, F., \& Van-Trung, D. 2001, in QSO Hosts and Their
Environments', ed. I. Marquez, preprint

\bibitem[2000]{lim00}
Lim, J., Leon, S., Combes, F., \& Dinh-V-Trung, D. 2000, ApJ, 545, L93

\bibitem[1996]{maloney96} 
Maloney, P. R., Hollenbach D. J., \& Tielens A. G. G. M. 1996, ApJ,
466, 561

\bibitem[2001]{marr01}
Marr, J. M., Taylor, G. B., \& Crawford, F. III 2001, ApJ, 550, 160

\bibitem[1999]{martel99}
Martel, A. R., Baum, S. A., Sparks, W. B., et al. 1999, ApJS, 122, 81

\bibitem[1988]{mirabel88}
Mirabel, I. F., \& Sanders, D. B. 1988, ApJ, 335, 104

\bibitem[1989]{mirabel89a}
Mirabel, I. F., 1989a, ApJ, 340, L13

\bibitem[1989]{mirabel89b}
Mirabel, I. F., Sanders, D. B., \& Kazes, I. 1989b, ApJ, 340, L9

\bibitem[1997]{morganti97}
Morganti, R., Oosterloo, T. A., Reynolds, J. E., Tadhunter, C.N., \&
Migenes, V. 1997, MNRAS, 284, 541

\bibitem[1999]{murgia99}
Murgia, M., Fanti, C., Fanti, R., et al. 1999, A\&A, 345, 769

\bibitem[1996]{odea96}
O'Dea, C. P., Stanghellini, C., Baum, S., \& Charlot, S. 1996, ApJ, 470,
806

\bibitem[1997]{odea97}
O'Dea, C. P., \& Baum, S. A. 1997, AJ, 113, 148

\bibitem[1999]{oosterloo99}
Oosterloo, T. A., Morganti, R., \& Sadler, E. 1999, PASA, 16, 28

\bibitem[1982]{orr82}
Orr, M. J. L., \& Browne, I. W. A. 1982, MNRAS, 200, 1067

\bibitem[1998]{owsianik98a}
Owsianik, I., \& Conway, J. E. 1998, A\&A, 337, 69

\bibitem[1998]{owsianik98b}
Owsianik, I., Conway, J. E., \& Polatidis, A. G. 1998, A\&A, 336, L37

\bibitem[1999]{owsianik99}
Owsianik, I., Conway, J. E., \& Polatidis, A. G. 1999, New Astronomy Reviews, 43, 669

\bibitem[1999]{parma99} 
Parma, P., Murgia, M., Morganti, R., et al. 1999, A\&A, 344, 7

\bibitem[1998]{peck98}
Peck, A. B., \& Taylor, G. B. 1998, ApJ, 502, L23

\bibitem[1999]{peck99}
Peck, A. B., Taylor, G. B., \& Conway, J. E. 1999, ApJ, 521, 103

\bibitem[2000]{peck00}
Peck, A. B., Taylor, G. B., Fassnacht, C. D., Readhead, A. C. S., \& Vermeulen,
R. C. 2000, ApJ, 534, 104

\bibitem[2001]{perlman01}
Perlman, E. S., Stocke, J. T., Conway, J. E., \& Reynolds, C. 2001,
AJ, 122, 536

\bibitem[2003]{polatidis03}
Polatidis, A. G., \& Conway, J. E. 2003, PASA, vol 20

\bibitem[1999]{polatidis99}
Polatidis, A., Wilkinson, P. N., Xu, W., et al. 1999, New Astronomy Reviews, 43, 657

\bibitem[1996]{readhead96}
Readhead, A. C. S., Taylor, G. B., Pearson, T. J., Wilkinson,
P. N. 1996, ApJ, 460, 634

\bibitem[2000]{sadler00}
Sadler, E. M., Oosterloo, T. A., Morganti, R., \& Karakas, A. 2000, AJ, 119,
1180

\bibitem[1995]{saikia95}
Saikia, D. J., Jeyakumar, S., Wiita, P. J., Sanghera, H. S., \&
Spencer, R. E. 1995, MNRAS, 276, 1215

\bibitem[1989]{spencer89} 
Spencer, R. E., McDowell, J. C., Charlesworth, M., et al. 1989, MNRAS, 240, 657

\bibitem[1997]{snellen97}
Snellen, I. A. G., PhD Thesis, Univ. of Leiden

\bibitem[2000]{snellen00}
Snellen, I. A. G., Schilizzi, R. T., Miley, G. K., de Bruyn, A. G.,
Bremer, M. N., \& R\"ottgering, H. J. A. 2000, MNRAS, 319, 445

\bibitem[1997]{stanghellini97}
Stanghellini C., Bondi M., Dallacasa D., et al.\ 1997, A\&A, 318, 376

\bibitem[1998]{stanghellini98}
Stanghellini, C., O'Dea, C. P., Dallacasa, D., et al. 1998, A\&AS, 131, 303

\bibitem[2001]{stanghellini01}
Stanghellini, C., Dallacasa, D., O'Dea, C. P., et al. 2001, A\&A, 377, 377

\bibitem[2000]{taylor00}
Taylor, G. B., Marr, J. M., Pearson, T. J. \& Readhead, A. C. S. 2000,
ApJ, 541, 112

\bibitem[1996]{taylor96}
Taylor, G. B. 1996, ApJ, 470, 394

\bibitem[2000]{tschager00}
Tschager, W., Schilizzi, R. T., R\"ottgering, H. J. A., Snellen,
I. A. G., \& Miley, G. K. 2000, A\&A, 360, 887

\bibitem[1986]{trinchieri86}
Trinchieri, G., Fabbiano, G., \& Canizares, C. R. 1986, ApJ, 310, 637

\bibitem[1984]{vanbreugel84}
van Breugel, W., Miley, G., \& Heckman, T., 1984, AJ, 89, 5

\bibitem[1989]{vangorkom89}
van Gorkom, J. H., Knapp, G. R., Ekers, R. D., et al. 1989, AJ, 97, 708

\bibitem[2000]{vanlangevelde00}
van Langevelde, H. J., Pihlstr\"om, Y. M., Conway, J. E., Jaffe,
 W., \& Schilizzi, R. T. 2000, A\&A, 354, L45

\bibitem[1999]{verdoes99}
Verdoes Kleijn, G. A., Baum, S. A., de Zeeuw, P. T., \& O'Dea,
C. P. 1999, AJ, 118, 2592

\bibitem[2002]{vermeulen02}
Vermeulen, R. C., Ros, E., Kellermann, K. I., et al. J. A. 2002, PASA 20, in press

\bibitem[2003]{vermeulen03}
Vermeulen, R. C., Pihlstr\"om Y.M., Tschager W., et al. 2003, A\&A, accepted

\bibitem[1995]{wiklind95}
Wiklind, T., Combes, F., \& Henkel, C. 1995, A\&A, 297, 643

\end{thebibliography}
\end{document}